\documentclass{article} 
\usepackage{iclr2021_conference,times}

\usepackage{amsmath,amsfonts,bm}

\def\eqref#1{equation~\ref{#1}}

\def\1{\bm{1}}

\DeclareMathAlphabet{\mathsfit}{\encodingdefault}{\sfdefault}{m}{sl}
\SetMathAlphabet{\mathsfit}{bold}{\encodingdefault}{\sfdefault}{bx}{n}

\usepackage{hyperref}
\usepackage{url}
\usepackage{graphicx}
\usepackage{adjustbox}
\usepackage{epstopdf} 

\title{LlamaRec-LKG-RAG: A Single-Pass, Learnable Knowledge Graph-RAG Framework for LLM-Based Ranking}

\author{Vahid Azizi, Fatemeh Koochaki\\
\texttt{\{va.azizi, fatmakoochaki\}}@gmail.com} 

\iclrfinalcopy 
\begin{document}

\maketitle


\begin{abstract}
Recent advances in Large Language Models (LLMs) have driven their adoption in recommender systems through Retrieval-Augmented Generation (RAG) frameworks. However, existing RAG approaches predominantly rely on flat, similarity-based retrieval that fails to leverage the rich relational structure inherent in user-item interactions. We introduce LlamaRec-LKG-RAG, a novel single-pass, end-to-end trainable framework that integrates personalized knowledge graph context into LLM-based recommendation ranking. Our approach extends the LlamaRec architecture by incorporating a lightweight user preference module that identifies salient relation paths within a heterogeneous knowledge graph constructed from user behavior and item metadata. These personalized subgraphs are seamlessly integrated into prompts for a fine-tuned Llama-2 model, enabling efficient and interpretable recommendations through a unified inference step. Comprehensive experiments on ML-100K and Amazon Beauty datasets demonstrate consistent improvements over LlamaRec across key ranking metrics (MRR, NDCG, Recall). LlamaRec-LKG-RAG demonstrates the critical value of structured reasoning in LLM-based recommendations and establishes a foundation for scalable, knowledge-aware personalization in next-generation recommender systems. Code is available at~\href{https://github.com/VahidAz/LlamaRec-LKG-RAG}{repository}.

\end{abstract}

\section{Introduction}
Large Language Models (LLMs) have achieved state-of-the-art performance across diverse natural language processing tasks, including summarization, translation, and question answering~\cite{brown2020language, zhang2022opt, touvron2023llama2openfoundation}. To enhance factual accuracy and domain adaptability, Retrieval-Augmented Generation (RAG) frameworks integrate LLMs with external retrieval mechanisms, enabling models to dynamically incorporate relevant contextual information during generation~\cite{lewis2020retrieval, izacard2021leveraging}. Despite their effectiveness, traditional RAG approaches predominantly rely on flat retrieval strategies that treat documents or passages as isolated units, ranking them solely based on vector similarity measures. This approach fundamentally overlooks the rich semantic relationships and structural dependencies between information elements, leading to fragmented retrievals and potentially incomplete responses for complex, multi-faceted queries. Graph-based Retrieval-Augmented Generation (GraphRAG) emerges as a promising solution to these limitations by representing information as structured graphs, where nodes capture entities and edges encode explicit semantic or relational dependencies~\cite{yasunaga2022qa, wang2024gnnrag, li2025grag, sun2023thinkongraph, chen2024thinkongraph, smith2025rgl, lee2025gfmrag}. This graph-structured representation enables sophisticated reasoning over interconnected information networks, facilitating multi-hop inference pathways and significantly improving the coherence and completeness of generated responses.

The rapid advancement of LLMs has positioned them as pivotal components in modern recommender systems, significantly enhancing personalization through robust representational capabilities, contextual reasoning, and generalization from limited data. Early methodologies primarily integrated LLMs as static modules, such as feature extractors or semantic enrichers, to refine user and item representations and process auxiliary information~\cite{wang2023survey, lin2023how}. However, these approaches often underutilize the interactive and generative potentials of LLMs. Recent research has shifted towards treating LLMs as active agents within the recommendation pipeline, enabling roles such as retrievers, rankers, or fully generative recommenders capable of producing ranked lists and explanations~\cite{cinar2023llm, chen2025llmagents}. This evolution facilitates advanced functionalities, including multi-turn dialogues, zero-shot personalization, and context-aware suggestions, steering recommender systems towards more conversational and explainable paradigms~\cite{liu2023chatrec, liu2024enhanced, chen2025llmagents, wang2023survey}.
Despite significant progress, challenges such as cold-start, data sparsity, and intent ambiguity remain central in recommender systems. RAG methods, particularly GraphRAG, have emerged as practical solutions to address these issues. By retrieving structured meta-knowledge from user-item graphs, knowledge graphs (KGs), or behavioral traces and integrating it with LLMs, GraphRAG enables fine-grained reasoning under sparse or noisy conditions~\cite{wang2025kragrec, peng2024graphragsurvey, han2025retrievalaugmentedgenerationgraphsgraphrag}. GraphRAG has been applied to diverse recommendation tasks, including graph-based recommendation~\cite{wang2025kragrec}, next-item prediction~\cite{cinar2023llm}, and conversational recommendation~\cite{liu2023chatrec}.


Integrating LLMs into recommendation systems introduces several challenges, including hallucinations, bias, limited controllability, lack of explainability, and high inference costs~\cite{zhao2023recommender, lin2023how, ma2023large, said2024explaining}. Recent research has explored scalable and interpretable solutions to mitigate these limitations. LlamaRec~\cite{yue2023llamarec} improves scalability with a two-stage architecture: a lightweight retriever first selects candidate items, followed by a fine-tuned Llama-2-7b~\cite{touvron2023llama2openfoundation} model for ranking. It employs a \textit{verbalizer}, a compact module that directly maps LLM logits to ranking scores in a single forward pass to reduce inference latency. To address hallucination and enhance reasoning in LLMs, Think-on-Graph (ToG)~\cite{sun2023thinkongraph} introduces an agent-based framework where the LLM performs multi-hop inference by interactively traversing a KG~\cite{sun2023thinkongraph}. This dynamic process grounds predictions in explicit, factual paths, improving transparency and interpretability. While LlamaRec offers efficiency, it underutilizes the LLM's reasoning capabilities and relies on the LLM's internal knowledge and data patterns. Conversely, ToG supports interpretable, structured reasoning but incurs latency due to iterative KG traversal, posing challenges for real-time, large-scale deployments for recommendation systems.

This work presents LlamaRec-LKG-RAG, an efficient and interpretable framework that integrates personalized information from KGs into LLM-based recommendation systems, aiming to enhance ranking performance. The proposed approach contributes two key innovations. First, we leverage the structured nature of recommendation data to construct an explicit KG comprising core entities, such as users, items, and ratings, alongside diverse relation types. This graph is further enriched with dataset-specific metadata, including item attributes (e.g., brand, category), to provide a richer semantic context. Second, we design a lightweight deep neural network called the user preference module that learns user-specific preferences over KG relation types. These preferences inform a relation-specific scoring function, enabling a precise, single-pass retrieval mechanism for extracting personalized subgraphs from the KG. The retrieved knowledge is then embedded into a carefully designed prompt template, combined with the user's historical interaction sequence and a candidate item set. This composite prompt is passed to a Llama-2-7b model, which performs the final ranking. The entire framework is trained end-to-end to optimize performance holistically, making LLM-based rankers integrate external personalized knowledge efficiently in a single pass and offering extensibility toward explainable recommendation systems. We evaluate our method against LlamaRec and report consistent improvements across standard recommendation benchmarks.

\section{Method}

\begin{figure}[]
  \centering
  \includegraphics[width=\textwidth]{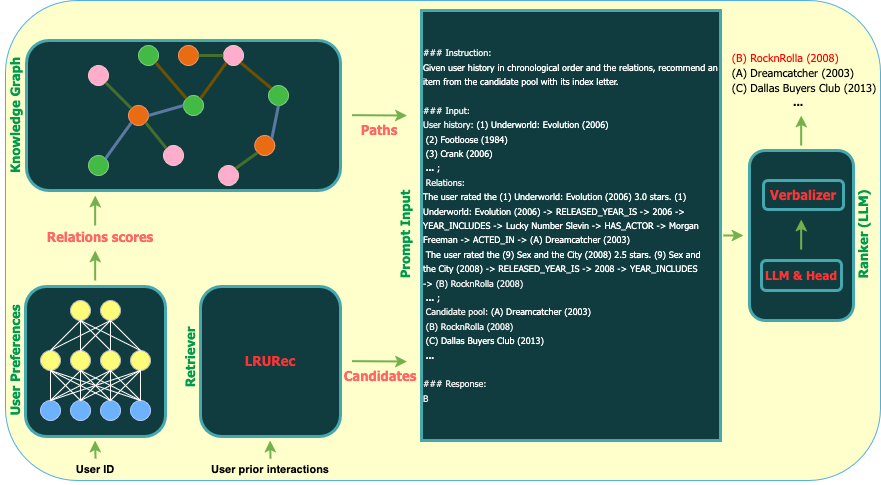}
  \caption{Overview of the LlamaRec-LKG-RAG framework. A sequential recommendation module (LRURec) generates candidate items from the user’s interaction history. A user preference module captures user-specific relation preferences to guide the retrieval of a personalized subgraph from a KG. The retrieved paths, user history, and candidate items are composed into a prompt, which is passed to a Llama-2-7b model for final ranking. The system is trained end-to-end to enable efficient and interpretable personalized ranking, augmented with user-specific knowledge extracted from the KG.}
  \label{model}
\end{figure}

We propose a novel and efficient methodology for extracting informative, user-personalized contextual information from KGs and seamlessly integrating this structured knowledge into LlamaRec's two-stage sequential recommendation framework to significantly enhance ranking performance. Our approach transforms the conventional LlamaRec architecture by incorporating a sophisticated knowledge graph reasoning component that enables personalized subgraph retrieval and integration. A comprehensive overview of the proposed LlamaRec-LKG-RAG architecture is presented in Fig.~\ref{model}, with detailed methodological exposition provided in the following sections.

\subsection{Retrieval Stage}
\label{retsec}
The candidate generation stage processes a chronologically ordered sequence of user interactions to produce a refined set of candidate items for subsequent ranking. We employ LRURec~\cite{yue2023linear, yue2023llamarec}, a computationally efficient sequential recommendation model built upon Linear Recurrent Units (LRUs) that maintains optimal performance while ensuring scalability. LRURec's architecture maintains a fixed-length hidden state representation, enabling efficient incremental updates and achieving low-latency inference capabilities essential for real-time recommendation deployment. The model is trained using an autoregressive objective that effectively captures complex user transition patterns and behavioral dynamics over time. During inference, LRURec generates refined item representations by computing dot products between predicted user state representations and learned item embeddings, enabling efficient similarity-based candidate identification. Given the extensive item catalog typical in recommendation scenarios, items are represented through unique identifier encodings to ensure computational tractability. We extract the top-K candidates (\textit{K=20}) from LRURec's output distribution, providing a focused candidate set for the subsequent LLM-based ranking stage while maintaining computational efficiency.

\subsection{Ranking Stage}
In the ranking stage following LlamaRec, we used the Llama-2-7b model. The LlamaRec authors created a prompt template~\cite{yue2023llamarec} that includes task instructions, the user’s recent interaction history, and candidate items as follows: 

\#\#\# Instruction: Given user history in chronological order, recommend an item from the candidate pool with its index letter.\\
\#\#\# Input: User history: {\textit{history}}; Candidate pool: {\textit{candidates}} \\
\#\#\# Response: {\textit{label}}

This format leverages item titles to help the model better understand user behavior and semantic signals. 
LlamaRec employs an innovative verbalizer-based ranking approach that assigns each candidate item a unique alphabetical index (A, B, C, etc.), transforming the ranking problem into a classification task. Rather than generating lengthy textual responses, the Llama-2-7b model efficiently identifies the most relevant item by outputting the corresponding index letter.
The ranking scores are directly extracted from the output logits associated with these index tokens, enabling the entire ranking process to be completed in a single forward pass. This approach significantly enhances computational efficiency and system scalability while circumventing the computational overhead and potential ambiguities inherent in traditional listwise or pairwise ranking methodologies. The verbalizer-based design ensures deterministic, interpretable outputs while maintaining the semantic reasoning capabilities of the underlying language model.

\subsection{LlamaRec-LKG-RAG}

Inspired by the ToG methodology, we have enhanced the LlamaRec framework by systematically integrating personalized contextual knowledge derived from structured knowledge graphs. This significantly improves the recommendation ranking performance. Our approach involves constructing a comprehensive, dataset-specific knowledge graph where users and items are the primary entities, connected through fundamental \textit{RATED} relations that capture explicit user feedback and interaction patterns. To further enrich the knowledge graph semantically, we augment it with additional metadata, such as item attributes (e.g., brand). This results in a heterogeneous knowledge graph that incorporates multiple types of nodes and relations.

To enable efficient personalized traversal of this KG, we design a lightweight user preference model that learns user-specific preferences for different relation types. The user preference model identifies and prioritizes each user's most relevant relation types based on patterns learned through interactions over time. This model consists of: (1) An embedding layer to encode each user into a dense vector representation. (2) A fully connected layer to transform these embeddings. (3) A final fully connected layer with softmax activation to predict a probability distribution over relation types in the KG. We adopt the LlamaRec prompting structure to format input for LLM inference with the following template:

\#\#\# Instruction: Given user history in chronological order and the relations, recommend an item from the candidate pool with its index letter.\\
\#\#\# Input: User history: {\textit{history}}; Relations: {\textit{Graph Paths}}; Candidate pool: {\textit{candidates}} \\
\#\#\# Response: {\textit{label}}

Here, \textit{history} and \textit{candidates} are placeholders for user interaction history and candidate items, respectively. \textit{Relations} correspond to personalized paths extracted from the KG based on user preferences. \textit{label} represents the target item for each training example, which is left empty during inference for prediction.
Our implementation focuses on retrieving semantic paths (information) between each history item and candidate item. Since multiple (or no) paths may exist, we extract the shortest path per pair with random tie-breaking to manage context size and remain within Llama-2's input limits. However, this still leads to up to \textit{K × K} potential relation paths, where \textit{K} is the number of items per set (see section~\ref{retsec}), potentially exceeding input constraints. To address this, we use the output of the user preference model to score each relation type, thereby assigning a relevance score to each path. A naive approach would sum the relation scores along a path. However, this can introduce bias if certain high-frequency relations dominate. To mitigate this, we adopt a (Term Frequency-Inverse Document Frequency) TF-IDF-inspired weighting scheme: the model's learned relation scores are scaled by the TF-IDF score of each relation in the context of the current query. Each path's final score reflects personalization and informativeness, and only the top-\textit{K}-scored paths are selected to be included in the LLM input context.

\section{Experiments}

\subsection{Dataset} We evaluated our approach using the ML-100K~\cite{10.1145/2827872} and Beauty~\cite{10.1145/2872427.2883037, 10.1145/2766462.2767755} datasets. The ML-100K dataset consists of 100,000 user-item interactions and is a standard benchmark for movie recommendation systems. In contrast, the Beauty dataset includes user reviews and interactions with beauty products collected from the Amazon platform. For preprocessing, we adhered to the protocol outlined in~\cite{yue2023llamarec, yang2023palrpersonalizationawarellms, 10.1145/3523227.3546770}. This protocol involves constructing input sequences in chronological order, removing items that lack metadata, and filtering out users and items with fewer than five interactions. The statistics for both datasets are summarized in Table~\ref{tab:stat}~\cite{yue2023llamarec}.

\begin{table*}[]
\centering
\caption{Summary statistics of the ML-100K and Beauty datasets, which include the number of users, items, interactions, and sequence lengths. Preprocessing follows the protocol in~\cite{yue2023llamarec}.}
\label{tab:stat}
\begin{tabular}{ccccc}
\hline
\textbf{Datasets} & \textbf{Users} & \textbf{Items} & \textbf{Interactions} & \textbf{Length} \\ \hline
\textbf{ML-100K} & 610 & 3650 & 100K & 147.99 \\
\textbf{Beauty} & 22332 & 12086 & 198K & 8.87 \\ \hline
\end{tabular}
\end{table*}

\subsection{Knowledge Graph (KG) Construction}
This section outlines the construction process of a knowledge graph (KG) for each dataset. We employed Neo4j~\cite{neo4j_homepage} for KG creation and utilized Cypher~\cite{neo4j_cypher_manual}, Neo4j’s declarative query language, to interact with the graph.

For the MovieLens dataset, the KG comprises four core entity types: users, movies (items), genres, and years. To enrich the graph with additional information, we extracted two more types of entities, directors and actors, from the IMDb website. The KG includes four primary types of relationships: (1) between users and movies, (2) between movies and actors, (3) between movies and directors, and (4) between movies and years. The user–movie relationship is represented as a unidirectional edge labeled \textit{RATED}, forming triples of the form (User, \textit{RATED}, Movie). In contrast, relationships involving movies and other entities (actors, directors, and years) are modeled bidirectionally to capture richer semantics. The relationship types are named to reflect their meaning, including:

\begin{itemize}
    \item (Movie, \textit{HAS\_ACTOR}, Actor) and (Actor, \textit{ACTED\_IN}, Movie)
    \item (Movie, \textit{DIRECTED\_BY}, Director) and (Director, \textit{IS\_THE\_DIRECTOR\_OF}, Movie)
    \item (Movie, \textit{RELEASED\_YEAR\_IS}, Year) and
    (Year, \textit{YEAR\_INCLUDES}, Movie)
\end{itemize}

\begin{figure}
  \centering
  \includegraphics[height=8cm]{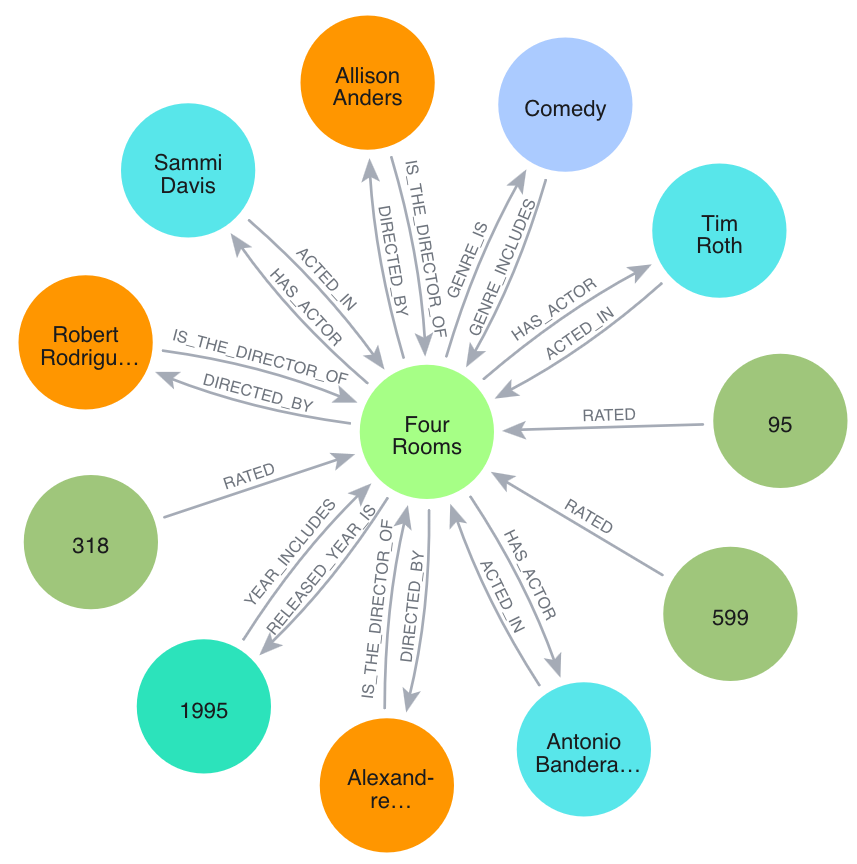}
  \caption{An illustrative subgraph of the knowledge graph constructed for the MovieLens dataset. The graph includes entity types such as users, movies, actors, directors, and years connected through semantically meaningful relationships. Unidirectional edges represent user ratings (e.g., (User, \textit{RATED}, Movie)), while bidirectional edges capture contextual associations, such as acting, directing, and release year.}
  \label{graph_ml}
\end{figure}

The final MovieLens KG contains 10,471 nodes and 130,002 edges. An illustrative subgraph of the MovieLens KG structure is presented in Fig.~\ref{graph_ml}.

The KG in the Beauty dataset comprises four primary entity types: users, items, product categories, and product brands. It includes five types of relationships among these entities. The user–item relationship is modeled as a unidirectional edge labeled \textit{RATED}, forming triples such as (User, \textit{RATED}, Item). Relationships between items, brands, and categories are modeled bidirectionally to capture semantic associations.

\begin{itemize}
    \item (Item, \textit{BRAND\_IS}, Brand) and (Brand, \textit{BRAND\_INCLUDES}, Item)
    
    \item (Item, \textit{CATEGORY\_IS}, Category) and (Category, \textit{CATEGORY\_INCLUDES}, Item)
\end{itemize}

Additionally, the dataset contains a set of unidirectional item–item relationships that reflect user behavior patterns, including:

\begin{itemize}
    \item (Item\_X, \textit{ALSO\_BOUGHT}, Item\_Y)
    
    \item (Item\_X, \textit{ALSO\_VIEWED}, Item\_Y)
    
    \item (Item\_X, \textit{BOUGHT\_TOGETHER}, Item\_Y)
    
    \item  (Item\_X, \textit{BUY\_AFTER\_VIEWING}, Item\_Y)
\end{itemize}

\begin{figure*}[]

  \centering
  \includegraphics[height=8cm]{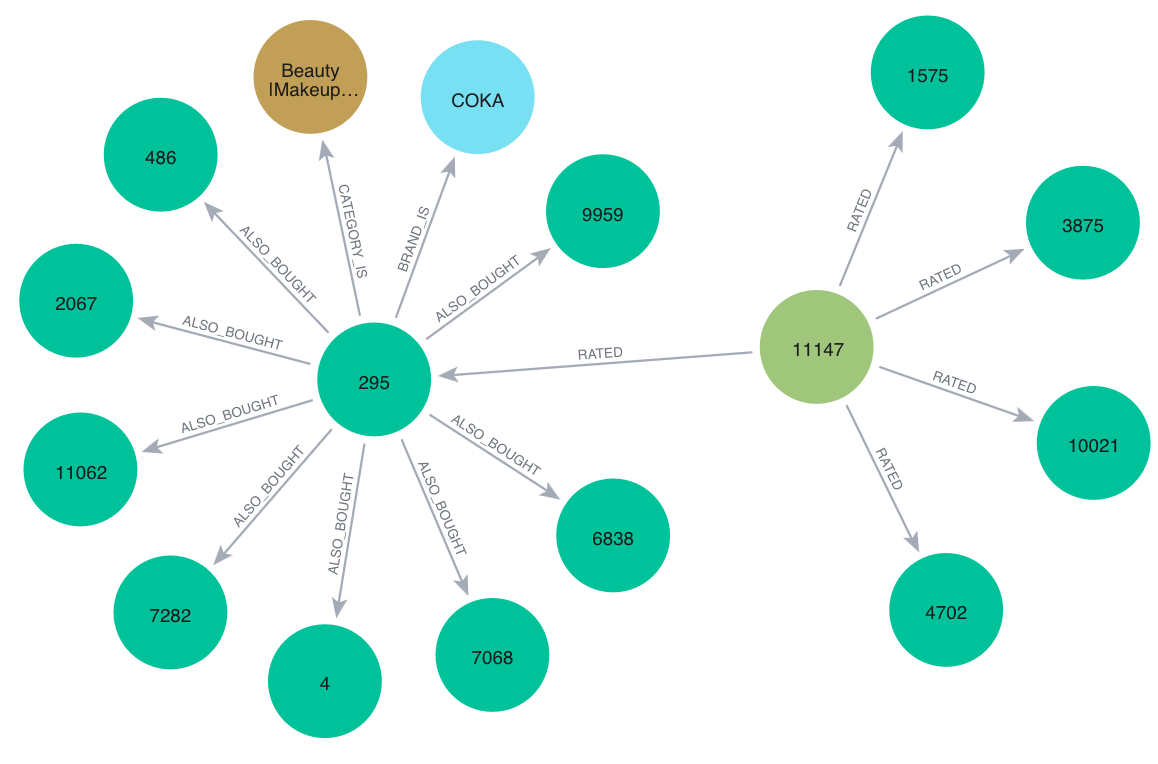}
  \caption{This is a representative subgraph of the knowledge graph constructed for the Beauty dataset. The graph includes users, items, brands, and product categories, connected via both unidirectional and bidirectional relationships. Item–item relations such as \textit{ALSO\_BOUGHT}, \textit{ALSO\_VIEWED}, and \textit{BOUGHT\_TOGETHER} capture user behavior patterns. After pruning infrequent or redundant relationships, the subgraph reflects the KG structure to enhance training and inference efficiency.}
  \label{fig:bt_subgraph}
\end{figure*}

Each type of relation captures the intuitive semantics suggested by its name; for example, \textit{ALSO\_BOUGHT} indicates that two items were purchased frequently together. 
Due to the large size of the Beauty KG, querying during training and inference can be computationally intensive. To mitigate this, we applied two pruning strategies. First, because of its low frequency, we remove the \textit{BUY\_AFTER\_VIEWING} relation. Second, for high-frequency relations such as \textit{ALSO\_VIEWED} and \textit{BOUGHT\_TOGETHER}, we retained only the most recent interaction per pair of items, preserving a single directional edge. Following pruning, the final KG contains 36,738 nodes and 543,088 relationships. A representative subgraph of Beauty KG is shown in Fig.~\ref{fig:bt_subgraph}.


\subsection{Baseline Methods and Evaluation}
In this study, we explore enhancing the ranking task with LLMs by efficiently integrating KGs. As our approach builds upon LlamaRec~\cite{yue2023llamarec}, we conduct a direct comparison exclusively against it. For evaluation, we adopt the leave-one-out strategy: for each user interaction sequence, the last item is reserved for testing, the second-to-last for validation, and the remaining items for training, following the protocol established in~\cite{yue2023llamarec}. To ensure a fair comparison, we employ the same evaluation metrics used in~\cite{yue2023llamarec}, including the mean reciprocal rank ($\text{MRR}@K$), the normalized discounted cumulative gain ($\text{NDCG}@K$) and the recall ($\text{Recall}@K$), where $K \in {1, 5, 10}$.

\subsection{Implementation} 
We employ LRURec~\cite{yue2023linear, yue2023llamarec} as the retriever module and train it separately for each dataset using the default hyperparameters provided in the original works~\cite{yue2023llamarec, yue2023linear}. For model quantization, we adopt QLoRA~\cite{10.5555/3666122.3666563} applied to Llama-2, using the following LoRA~\cite{hu2021loralowrankadaptationlarge} configuration: rank dimension = 8, $\alpha = 32$, dropout = 0.05, and learning rate = $1 \times 10^{-4}$. The target modules are the \textit{Q} and \textit{V} projection matrices. Consistent with LlamaRec~\cite{yue2023llamarec}, we fix the number of items retained for the user history and candidate set at 20. Incorporating KG information increases the textual input length. Specifically, the input text length is set to 2,286 tokens for the MovieLens dataset and 2,536 tokens for the Beauty dataset. The maximum title length is capped at 32 tokens for MovieLens and 10 tokens for Beauty. The architecture of the user preference module is consistent across datasets. It consists of an embedding layer initialized using Xavier uniform initialization, followed by layer normalization~\cite{ba2016layernormalization}, a dropout layer (dropout rate = 0.2), a fully connected layer, and a softmax activation for classification. For MovieLens, the user embedding dimension is set to 128, with four output classes corresponding to KG relation types: \textit{GENRE\_IS}, \textit{RELEASED\_YEAR\_IS}, \textit{DIRECTED\_BY}, and \textit{HAS\_ACTOR}. For Beauty, the user embedding dimension is 512, with five output classes reflecting its KG relation types: \textit{BRAND\_IS}, \textit{CATEGORY\_IS}, \textit{ALSO\_BOUGHT}, \textit{ALSO\_VIEWED}, and \textit{BOUGHT\_TOGETHER}. All models are trained for one epoch and validated every 100 iterations. Early stopping is applied with a patience parameter of 20 iterations. The model checkpoint with the best validation performance is evaluated on the test set.

\subsection{Results}
The performance comparison between our model (LlamaRec-LKG-RAG) and LlamaRec is presented in Table~\ref{tblres}, reporting results for $\text{MRR}@K$, $\text{NDCG}@K$, and $\text{Recall}@K$ at $K \in {1, 5, 10}$. Our model consistently outperforms LlamaRec across most metrics, although the degree of improvement varies between the two datasets. On the MovieLens dataset, our model demonstrates clear superiority across all evaluation metrics, with particularly notable performance gains. In contrast, the improvements in the Beauty dataset are more modest. At $K = 5$, for instance, both models yield nearly identical scores for NDCG and Recall. However, our model achieves approximately a $1\%$ gain in MRR, which is especially meaningful since MRR emphasizes the rank of the first relevant item, aligning well with our ranking-centric objective. The relatively minor performance margin on the Beauty dataset may be attributed to its sparser interaction patterns, which limit the effectiveness of the user preference module. With fewer interactions per user, accurately modeling user behavior becomes more challenging, thereby reducing the impact of KG integration.

\begin{table}
\caption{Performance comparison between LlamaRec-LKG-RAG and LlamaRec on the MovieLens and Beauty datasets. Results are reported for $\text{MRR}@K$, $\text{NDCG}@K$, and $\text{Recall}@K$ at $K \in {1, 5, 10}$. LlamaRec-LKG-RAG consistently outperforms the baseline, with notable gains on the MovieLens dataset, underscoring the effectiveness of incorporating KG information for the ranking task.}
\label{tblres}
\begin{tabular}{ccc|cc}
 & \multicolumn{2}{c|}{\textbf{ML-100K}} & \multicolumn{2}{c}{\textbf{Beauty}} \\
 & \textbf{LlamaRec-LKG-RAG} & \textbf{LlamaRec} & \textbf{LlamaRec-LKG-RAG} & \textbf{LlamaRec} \\ \hline
\textbf{MRR@1} & \textbf{0.0262 ($\sim$22.5\% $\uparrow$)} & 0.0214 & \textbf{0.0225 ($\sim$5\% $\uparrow$)} & 0.0214 \\
\textbf{NDCG@1} & \textbf{0.0262 ($\sim$22.5\% $\uparrow$)} & 0.0214 & \textbf{0.0225 ($\sim$5\% $\uparrow$)} & 0.0214 \\
\textbf{Recall@1} & \textbf{0.0262 ($\sim$22.5\% $\uparrow$)} & 0.0214 & \textbf{0.0225 ($\sim$5\% $\uparrow$)} & 0.0214 \\ \hline
\textbf{MRR@5} & \textbf{0.0417 ($\sim$9\% $\uparrow$)} & 0.0383 & \textbf{0.0349 ($\sim$1\% $\uparrow$)} & 0.0346 \\
\textbf{NDCG@5} & \textbf{0.0499 ($\sim$7\% $\uparrow$)} & 0.0467 & \textbf{0.0409 ($\sim$equal)} & 0.0407 \\
\textbf{Recall@5} & \textbf{0.0754($\sim$4.5\% $\uparrow$)} & 0.0721 & \textbf{0.0593($\sim$equal)} & 0.0594 \\ \hline
\textbf{MRR@10} & \textbf{0.0462 ($\sim$7\% $\uparrow$)} & 0.0431 & \textbf{0.0386 ($\sim$1.5\% $\uparrow$)} & 0.0380 \\
\textbf{NDCG@10} & \textbf{0.0609 ($\sim$5\% $\uparrow$)} & 0.0580 & \textbf{0.0498 ($\sim$1.5\% $\uparrow$)} & 0.0491 \\
\textbf{Recall@10} & \textbf{0.1098 ($\sim$3\% $\uparrow$)} & 0.1065 & \textbf{0.0868($\sim$1.5\% $\uparrow$)} & 0.0855 \\ \hline
\end{tabular}
\end{table}

\begin{table*}[h]
\centering
\caption{Ablation study on the MovieLens dataset evaluating the impact of incorporating shortest paths between historical and candidate items as contextual input (\textit{LlamaRec-KG-RAG}). The results show that introducing KG paths without a relevance filtering mechanism degrades performance, highlighting the critical role of the user preference module in selecting and integrating meaningful, user-personalized KG context. Bold values indicate the lowest performance scores across models.}
\label{tblresstudy}
\begin{tabular}{cccc}
 & \multicolumn{3}{c}{\textbf{ML-100K}} \\
 & \textbf{LlamaRec-LKG-RAG} & \textbf{LlamaRec-KG-RAG} & \multicolumn{1}{l}{\textbf{LlamaRec}} \\ \hline
\textbf{MRR@1} & 0.0262 & \textbf{0.0180} & 0.0214 \\
\textbf{NDCG@1} & 0.0262 & \textbf{0.0180} & 0.0214 \\
\textbf{Recall@1} & 0.0262 & \textbf{0.0180} & 0.0214 \\ \hline
\textbf{MRR@5} & 0.0417 & \textbf{0.0324} & 0.0383 \\
\textbf{NDCG@5} & 0.0499 & \textbf{0.0382} & 0.0467 \\
\textbf{Recall@5} & 0.0754 & \textbf{0.0557} & 0.0721 \\ \hline
\textbf{MRR@10} & 0.0462 & \textbf{0.0380} & 0.0431 \\
\textbf{NDCG@10} & 0.0609 & \textbf{0.0517} & 0.0580 \\
\textbf{Recall@10} & 0.1098 & \textbf{0.0967} & 0.1065 \\ \hline
\end{tabular}
\end{table*}

\begin{figure*}[]
  \centering
  \includegraphics[width=\linewidth]{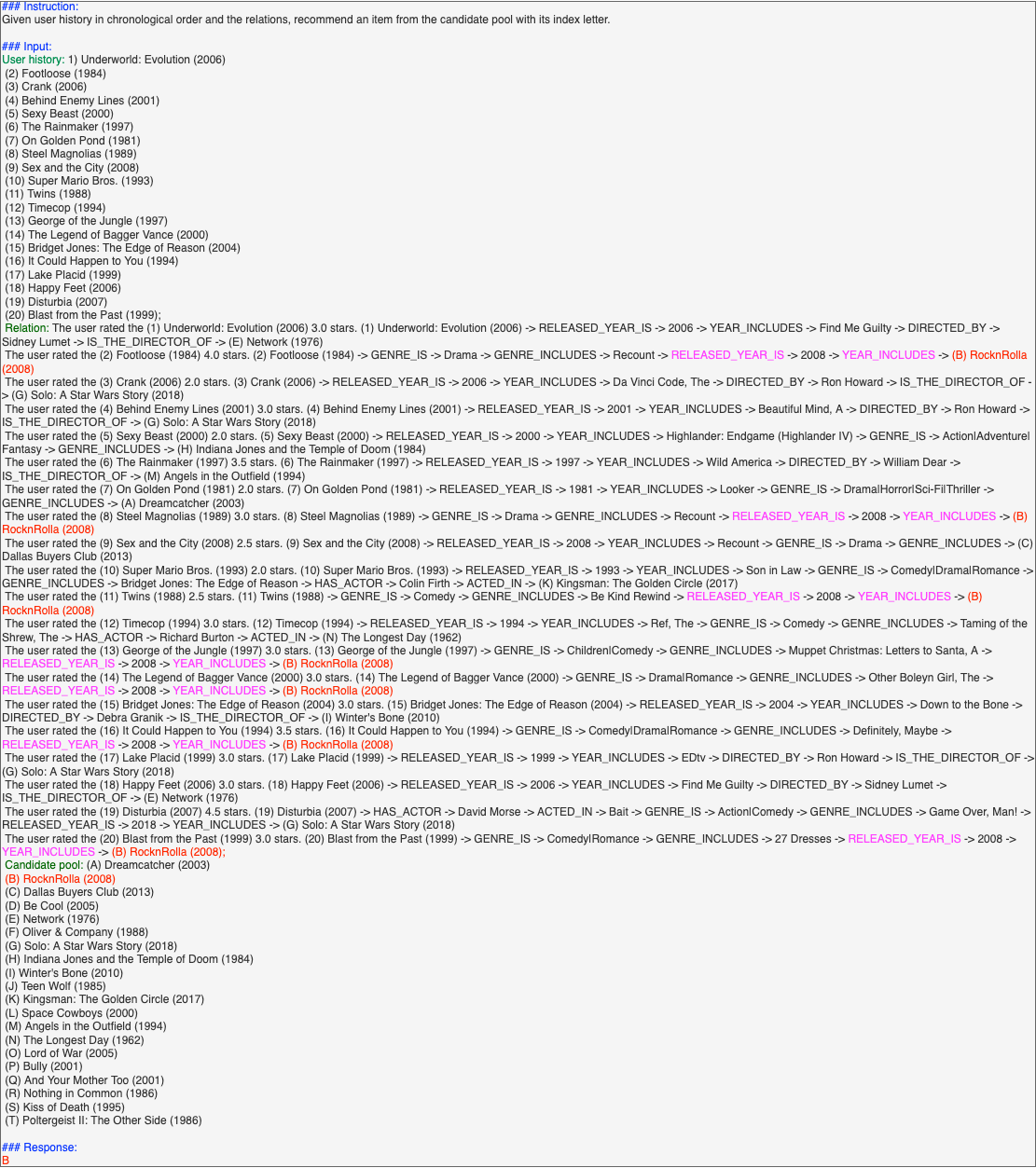}
  \caption{An illustrative example of a model query enriched with KG context selected using the user preference module. The correct answer \textit{(B) RocknRolla (2008)} appears in multiple relation paths, particularly those involving the \textit{RELEASED\_YEAR\_IS} relation. These consistent and semantically meaningful connections help the LLM reason effectively and make accurate predictions aligned with the user's inferred preferences.}
  \label{samplewkg}
\end{figure*}

\begin{figure}
  \centering
  \includegraphics[width=\linewidth]{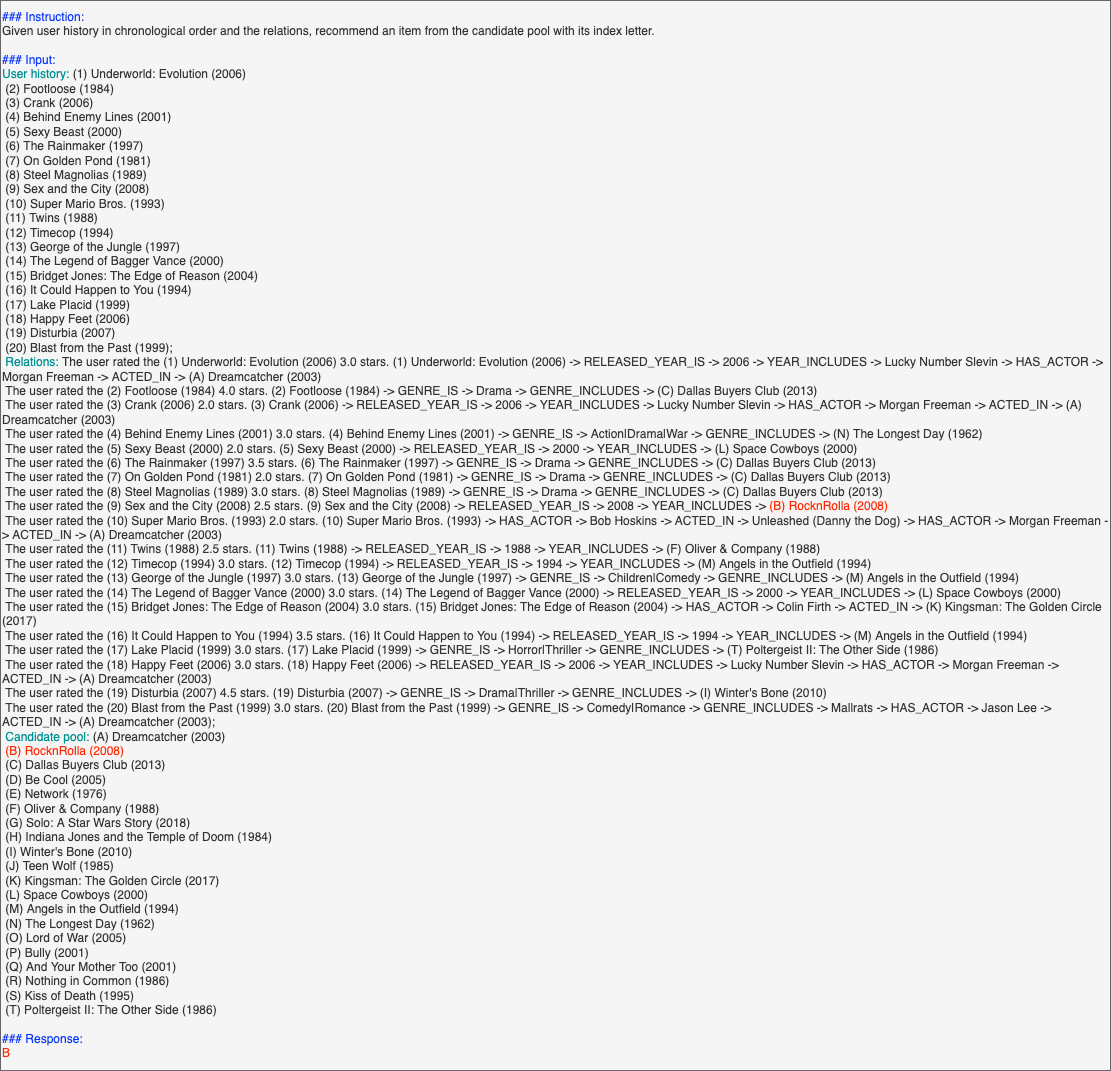}
  \caption{A prompt generated using KG context from \textit{LlamaRec-KG-RAG} without the user preference module. The included relations between items are arbitrary and lack user-specific relevance, resulting in inconsistent and noisy contextual information. This unfiltered input hinders the model’s reasoning ability and leads to incorrect predictions.}
  \label{samplewokg}
\end{figure}

\subsection{Ablation Study} 
We conducted an additional experiment on the MovieLens dataset to evaluate the impact of incorporating KG context without utilizing the user preference module. In this variant, we extracted the shortest paths between each historical and candidate item pair, randomly selecting one in cases of multiple shortest paths, and included them as contextual input, denoted as \textit{LlamaRec-KG-RAG}. The results, shown in Table~\ref{tblresstudy}, indicate that injecting information without a filtering or relevance mechanism leads to a decline in model performance. This finding underscores the importance of integrating KG information selectively. In particular, it highlights the value of the user preference module, which captures user-specific behavior to identify the most relevant information. By personalizing the contextual input, the module ensures that only semantically meaningful and user-aligned information are incorporated, ultimately enhancing the effectiveness of the ranking model. Fig.~\ref{samplewkg} illustrates that the model prioritizes certain relation types for a given user, such as release year. In this example, the correct answer (Answer B) appears as an entity in multiple paths involving the year relation. Equipped with this structured context, the LLM can reason over these relations and infer that the user is most likely interested in the movie \textit{RocknRolla (2008)}.
In contrast, Fig.~\ref{samplewokg} presents the prompt for the same user, generated by \textit{LlamaRec-KG-RAG}, which does not utilize the user preference module. As shown, the selected paths between items do not follow any coherent or user-aligned pattern. This lack of structure introduces noise and ambiguity, making it difficult for the model to perform practical reasoning. Consequently, the model is unable to identify the correct answer.


\subsection{Discussion}
Several promising directions and configurations warrant further exploration. Below, we outline some of the most noteworthy avenues for future work.

This study utilized the Llama-2 model, which has a constrained context length of 4096 tokens and limited reasoning capacity. We conducted exploratory experiments with larger models, such as ChatGPT~\cite{openai2025chatgpt} and Llama-2-70b~\cite{touvron2023llama2openfoundation}, examining both configurations, with and without our proposed user preference module that filters paths versus using all \textit{K×K} paths. Our preliminary results indicate that larger models perform well in both configurations. This suggests that models with longer input capacities and enhanced reasoning abilities may not require explicit information filtering. However, this approach poses two significant challenges: (1) extracting all \textit{K×K} paths from the knowledge graph is computationally intensive and slow for real-time applications like recommendation systems, and (2) processing extended input sequences greatly increases inference costs.

One key benefit of our approach is its ability to generate explanations. We instructed LLMs, specifically ChatGPT, Llama-2-7b, and Llama-2-70b, to rank items and provide reasoning for their choices. Notably, all models, particularly the larger ones, produced coherent and plausible explanations. This experiment highlights a promising direction for future research on explainable recommendation systems leveraging LLMs and KGs~\cite{a11090137, said2024explaining}.

As mentioned, we utilized raw relation paths from the knowledge graph without verbalizing~\cite{han2025retrievalaugmentedgenerationgraphsgraphrag} them. However, converting KG triples and paths into natural language has been explored and presents an intriguing avenue for future work. This verbalization could enhance the language model's understanding of the graph-based context, offering a rich area for further investigation.

We used a scoring strategy based on TF-IDF for path selection and ranking. Before implementing this approach, we tested various alternative heuristics, but they produced inconsistent results across our datasets. In contrast, TF-IDF consistently enhanced performance in both scenarios. Filtering strategies, such as excluding paths linked to low item ratings, can also affect performance. Conducting a systematic study of these filtering and ranking strategies is a promising avenue for future research.

In LlamaRec, item titles were utilized to guide ranking, drawing on the language model's internal knowledge. However, in specific datasets, this internal knowledge may be incomplete, making the use of item IDs a more suitable choice. In our experiments, our model (primarily relying on graph-based context) outperformed LlamaRec when using IDs instead of titles. This finding underscores the effectiveness of our approach when textual information is unreliable or unavailable, suggesting that further investigation in this area is warranted.

We augmented the KG with semantic metadata, such as brand information, but we did not include user-specific metadata in this study. Linking users to relevant nodes could incorporate user attributes, like age group in the MovieLens dataset, creating richer paths and potentially enhancing personalization. The effect of such user-level enrichment is an open question that warrants further research.

We extracted the paths between historical items and candidate items generated by LRURec, which outperformed traditional collaborative filtering methods. However, we did not examine how the quality of the retriever model influences overall performance. Since path extraction relies heavily on retrieved items, studying this interaction represents a valuable direction for future work.

Recent advancements in LLMs have significantly improved their reasoning capabilities, particularly through reinforcement learning-based fine-tuning techniques such as Reinforcement Learning from Human Feedback (RLHF). These techniques enable LLMs to perform multi-step reasoning, align with human intent, and generate contextually relevant and logically consistent outputs. Notably, these capabilities are starting to influence the design of next-generation recommendation systems. By integrating LLMs fine-tuned for reasoning, modern recommender frameworks can progress beyond static preference modeling to provide more interactive, context-aware, and goal-directed recommendations. Further integration and training of such frameworks using RLHF and Reinforcement Learning from AI Feedback (RLAIF) techniques could be a promising avenue, especially in live settings.

Finally, we constructed the KG offline before running our model. However, our framework is sequential, meaning temporal consistency must be maintained. Adding metadata, such as director information in the MovieLens dataset or brand details in the Beauty dataset, could inadvertently introduce paths containing future information, thereby violating causal constraints. To address this, a key direction for future work is to build the KG dynamically, filtering out nodes or edges with timestamps that exceed the current point in time. This approach would more accurately simulate real-world deployment and provide valuable insights into temporal generalization.

\section{Conclusion}
In this paper, we presented an efficient framework for integrating knowledge graphs (KGs) into LLM-based recommendation systems, with a focus on enhancing ranking performance. Our contributions are twofold: (1) we introduce the use of KGs as structured, contextual input to support LLM-based item ranking, and (2) we propose a lightweight neural network that models personalized user preferences to guide the extraction of relevant subgraphs from the KG. Building on LlamaRec, we augmented its two-stage recommendation pipeline with these components to enable efficient, interpretable, and personalized ranking. Extensive experiments on the MovieLens and Amazon Beauty datasets demonstrate that our approach consistently outperforms the baseline across standard recommendation metrics. These findings underscore the effectiveness of combining structured, user-aligned knowledge with LLMs and suggest promising directions for future work in scalable, knowledge-aware, and explainable recommendation systems.



\bibliography{iclr2021_conference}
\bibliographystyle{iclr2021_conference}


\end{document}